\documentstyle[preprint,aps]{revtex}
\begin{document}
\draft
\title{Remarks on the $\Delta^+$ mass}
\author{Ron Workman}
\address{Department of Physics, Virginia Polytechnic Institute and State
University, Blacksburg, VA 24061}
\date{\today}
\maketitle
\begin{abstract}

We question the reliability of current estimates for the 
$\Delta^+$ mass. We explain why the standard 
value, determined from pion photoproduction, should not be used
in conjunction with present estimates for the $\Delta^{++}$ and
$\Delta^0$ masses, determined from pion-nucleon scattering. 
The resultant mass splittings are internally inconsistent. 
We also briefly comment on discrepancies 
found when `theoretical' and `experimental' values for the 
$\Delta$ masses are compared. 

\end{abstract}

 
\narrowtext
\section{INTRODUCTION}
 
What is the $\Delta (1232)$ mass? As the name implies, 1232 MeV is a 
reasonable estimate for the `average' mass. The $\Delta^0$ and 
$\Delta^{++}$ masses, 1233.6$\pm$0.5 MeV and 1230.9$\pm$0.3 MeV 
respectively\cite{koch,pdg96}, have been estimated from pion-nucleon 
scattering, and the $\Delta^+$ mass (1234.9$\pm$1.4 MeV) has been 
extracted\cite{miro} from pion photoproduction data. 
Taken together, the $\Delta^+$ mass seems inconsistent
with estimates for the $\Delta^0$ and $\Delta^{++}$ masses. To be more
specific, the resultant splittings do not fit neatly into
the pattern exhibited by other members of the baryon octet and
decuplet\cite{cut,luty}. 

While the above result of Ref.\cite{miro} is usually quoted\cite{pdg96} 
for the $\Delta^+$ mass, most estimates have resulted in values which are 
considerably lower and closer to theoretical estimates. In the following,
we first explain why the $\Delta^+$ mass of Ref.\cite{miro} should not be
used in conjunction with more recent determinations of the $\Delta^0$ and
$\Delta^{++}$ masses. We then 
emphasize that similar problems arise in any analysis starting from 
multipoles determined in a separate analysis. Finally, we note that pole 
parameters may be in better agreement with theoretical estimates. The
splitting found in $\Delta$ pole masses\cite{com} is given as an example.  

\section{THE $\Delta (1232)$ AND THE $\Delta (1236)$}     

Values for the $\Delta^+$ mass and pole parameters, found in 
Ref.\cite{miro}, were based upon a multipole analysis performed in
Ref.\cite{miro2}. In that analysis, the multipoles with $L\le 1$ were
constrained, via Watson's theorem\cite{watson}, to have the phases
of the corresponding pion nucleon partial waves. 
Phases were taken from the CBC\cite{CBC}
and CERN\cite{CERN} analyses, and the results were compared. 
The authors concluded that the CERN phases were more consistent with the
data base being fitted. The result of this choice can be seen in Table~I of 
Ref.\cite{miro}. The real part of the $M_{1+}^{3/2}$ multipole passes
through zero for a photon energy near 350 MeV. However, in more recent 
analyses\cite{han}, the cross over point is closer to 340 MeV. 
In fact, at the time of the CERN analyses, the $P_{33}$ resonance was 
named the $\Delta (1236)$\cite{PDG},  
the CERN value\cite{CERN} being 1235.8 MeV. (Note that the 
$\Delta^+$ mass (1234.9$\pm$1.4 MeV) found in Ref.\cite{miro} is 
actually consistent with the CERN value.) Clearly the results of this
analysis\cite{miro} cannot be combined with those found in more recent 
pion nucleon analyses\cite{kh,cmb,vpi}. A 4 MeV shift in the expected 
`average mass' overwhelms the splitting of charge states. 

Since the use of Watson's theorem determines the phase behavior exhibited by 
photoproduction multipoles, we expect to find a $\Delta^+$ mass between the
$\Delta^0$ and $\Delta^{++}$ masses. This has generally been the result of
analyses using more modern pion nucleon phases\cite{pdg96}. This result 
appears reasonable, 
based on the splitting seen in the $\Sigma$ and $\Sigma^*$ 
states. The most model-independent determination would require sufficient
observables to drop the Watson's theorem constraint. Another approach would
be to allow small shifts in the $M^{3/2}_{1+}$ and $E^{3/2}_{1+}$ phases.
This was attempted in Ref.\cite{bd}, utilizing the CBC phases\cite{CBC}.
The estimated $\Delta^+$ mass was 1231.8 MeV, lying about halfway between
the $\Delta^0$ and $\Delta^{++}$ masses estimated in the CBC 
analysis\cite{CBC}. If an approximate 
equal-spacing rule holds for the $\Delta$ states, it will be very difficult
to distinguish between results with/without the Watson's theorem 
constraint. 

\section{MASS FORMULAE AND MASS SPLITTING}

While the above arguments result in a more satisfying set of
$\Delta$ masses, we should remind the reader of some problems which
persist. One recent paper\cite{bern}, 
relevant to these issues, has redetermined
the $\Delta^0$ and $\Delta^{++}$ resonance and pole parameters from a
fit to the Pedroni total cross sections\cite{pedroni}. 

The resulting $\Delta^0 - \Delta^{++}$ mass difference was 
2.25$\pm$0.68 MeV, a value consistent (within errors) with the original
Pedroni determination (2.7$\pm$0.3 MeV)\cite{pedroni}. 
A rough estimate of the expected $\Delta$ splitting,
in terms of the $n-p$ mass difference, was given as
\begin{equation}
\Delta^0 - \Delta^{++} = 2(n-p) ,
\end{equation}
which again seems reasonable in light of the mass relation\cite{luty}
\begin{equation}
\Delta^0 - \Delta^+ = n-p  .
\end{equation}
The predicted splitting is then about 2.6 MeV, which agrees with the 
values found in Refs\cite{bern,pedroni}. 
However, the more formally correct mass relation\cite{luty} 
for this splitting  
\begin{equation}
\Delta^0 - \Delta^{++} = 2(n-p) - \left( \Sigma^+ - 2\Sigma^0 
 + \Sigma^- \right)  ,
\end{equation}
agrees with the previous estimate only in the case of equally 
spaced charge states. Inserting experimental values\cite{pdg96}, 
one obtains a 0.9 MeV
splitting between the $\Delta^0$ and $\Delta^{++}$. 

In a fit to the octet and decuplet masses, 
Cutkosky\cite{cut} found a $\Delta^0 - \Delta^{++}$ splitting of 
1.53 MeV. This dropped to 0.81 MeV
when the experimental $\Delta$ masses were excluded from the fit.
(No attempt was made to fit the $\Delta^+$ mass of Ref.\cite{miro}.) 
These values are actually closer to the splitting between pole masses 
(0.40$\pm$0.57 MeV) found in Ref.\cite{bern}. A similarly 
small splitting (0.8 MeV) is found in the Virginia Tech analysis of 
pion nucleon scattering data\cite{raa}.

Finally, we note that the fit by Cutkosky\cite{cut} is constructed to 
satisfy the mass relations in Eqs.~(2) and (3). Since the predicted
$\Delta^0 - \Delta^{++}$ splitting is {\it less} than the 
$\Delta^0 - \Delta^+$ splitting, the $\Delta^+$ mass falls 
below the $\Delta^{++}$. This possibility provides motivation for
studies where the Watson's theorem constraint is relaxed. 

\section{SUMMARY AND CONCLUSIONS}
\label{sec:sum}

In this paper, we have seen that the standard value for the $\Delta^+$
mass\cite{miro} is inconsistent with $\Delta^0$ and $\Delta^{++}$ masses 
found in more recent pion nucleon analyses. The problem can be traced to
the use of outdated phase information. In fact, the use of Watson's theorem
always biases the masses found in pion photoproduction analyses. While
a $\Delta^+$ mass midway between the $\Delta^0$ and $\Delta^{++}$ masses
appears more reasonable, and is supported by the study made in Ref.\cite{bd},
the fit made by Cutkosky\cite{cut} does not confirm this simpler picture. 
 
The $\Delta^+$ pole position\cite{han,asw} is consistent with values  
found for mixed charges 
from pion nucleon scattering\cite{pdg96}. However, since the 
splitting in pole positions is considerably smaller than that found for the
resonance masses, an accurate determination of the splitting 
($\Delta^0 - \Delta^+$) will be difficult. The apparent agreement between 
pole mass splittings and mass formulae is suggestive, but could be 
accidental given the uncertainties associated with both the pole splitting
and the mass formula in Eq.~(3). 
 
\acknowledgments
 
We thank R.A. Arndt and G. Keaton for helpful discussions. 
This work was supported in part by a U.S. Department of
Energy Grant DE-FG05-88ER40454.


\begin{references}
\bibitem{com}
By pole mass we mean the real part ($m$) of the pole
position ($m-i\Gamma /2$). For the case of mixed charges,
the $\Delta$ pole is found at 1210$-$i50 MeV.
\end{references}
\end{document}